\documentclass[12pt]{article}
\usepackage{graphicx}
\usepackage{lineno}

\newcommand\pubnumber{}
\newcommand\pubdate{\today}

\def\institute{Deutsches Elektronen-Synchrotron DESY, Notkestrasse 85, D-22607 Hamburg, GERMANY}
\def\support{\footnote{Work supported by the Helmholtz association.}}

\def\Title#1{\begin{center} {\Large #1 } \end{center}}
\def\Author#1{\begin{center}{ \sc #1} \end{center}}
\def\Address#1{\begin{center}{ \it #1} \end{center}}

\newcommand\pubblock{\rightline{\begin{tabular}{l} \pubnumber\\
         \pubdate  \end{tabular}}}
\newenvironment{Abstract}{\begin{quotation}  }{\end{quotation}}
\newenvironment{Presented}{\begin{quotation} \begin{center} 
             PRESENTED AT\end{center}\bigskip 
      \begin{center}\begin{large}}{\end{large}\end{center} \end{quotation}}

\def\beq{\begin{equation}}
\def\eeq#1{\label{#1}\end{equation}}
\def\eeqn{\end{equation}}

\def\beqa{\begin{eqnarray}}
\def\eeqa#1{\label{#1}\end{eqnarray}}
\def\eeqan{\end{eqnarray}}

\let\bar=\overbar

\def\Dslash{\not{\hbox{\kern-4pt $D$}}}
\def\dslash{\not{\hbox{\kern-2pt $\del$}}}

\def\msb{{\bar{\ssstyle M \kern -1pt S}}}

\newcommand{\fbi}{fb$^{-1}$}

\newcommand*{\TeV}{\ifmmode {\mathrm{\ Te\kern -0.1em V}}\else
                   \textrm{Te\kern -0.1em V}\fi}%
\newcommand*{\GeV}{\ifmmode {\mathrm{\ Ge\kern -0.1em V}}\else
                   \textrm{Ge\kern -0.1em V}\fi}%
\newcommand*{\MeV}{\ifmmode {\mathrm{\ Me\kern -0.1em V}}\else
                   \textrm{Me\kern -0.1em V}\fi}%
\newcommand*{\keV}{\ifmmode {\mathrm{\ ke\kern -0.1em V}}\else
                   \textrm{ke\kern -0.1em V}\fi}%
\newcommand*{\eV}{\ifmmode  {\mathrm{\ e\kern -0.1em V}}\else
                   \textrm{e\kern -0.1em V}\fi}%
\newcommand{\MET}{\ensuremath{E^{\textrm{{\scriptsize miss}}}_{\textrm{{\scriptsize T}}}}}
\newcommand{\HT}{\ensuremath{H_{\textrm{{\scriptsize T}}}}}

\begin{document}
\begin{titlepage}
\pubblock

\vfill
\Title{Why stop at two tops?\\
       Search for exotic production of top quarks in final states with same-sign leptons and $b$-jets at 13 TeV }
\vfill
\Author{C\'{e}cile Deterre on behalf of the ATLAS Collaboration\support}
\Address{\institute}
\vfill
\begin{Abstract}
An analysis is presented of events containing jets including at
least one $b$-tagged jet, sizable missing transverse momentum, and at
least two charged leptons including a pair of the same electric
charge, with the scalar sum of the jet and lepton transverse momenta
being large. Standard Model processes rarely produce these final
states, but several models of physics beyond the Standard Model
predict an enhanced production rate of such events. Specific models
with this feature are considered here: vector-like $T$, $B$, and $T_{5/3}$
quark pair production, and four top quark production under three
scenarios (Standard Model, contact interaction, and extra-dimensions). 
A data sample of 3.2 \fbi\ of $pp$ collisions at a center-of-mass 
energy of $\sqrt{s}$=13 \TeV\ recorded by the ATLAS detector at the
Large Hadron Collider is used in this analysis. Several signal
regions are defined, in which the consistency between the data yield
and the background-only hypothesis is checked, and 95\% confidence
level limits are set on various signal models.
The focus here is on models yielding signatures with four top quarks.
\end{Abstract}
\vfill
\begin{Presented}
$9^{th}$ International Workshop on Top Quark Physics\\
Olomouc, Czech Republic,  September 19--23, 2016
\end{Presented}
\vfill
\end{titlepage}
\def\thefootnote{\fnsymbol{footnote}}
\setcounter{footnote}{0}

\section{Introduction}

\begin{figure}[htb]
\centering
\includegraphics[width=0.27\linewidth]{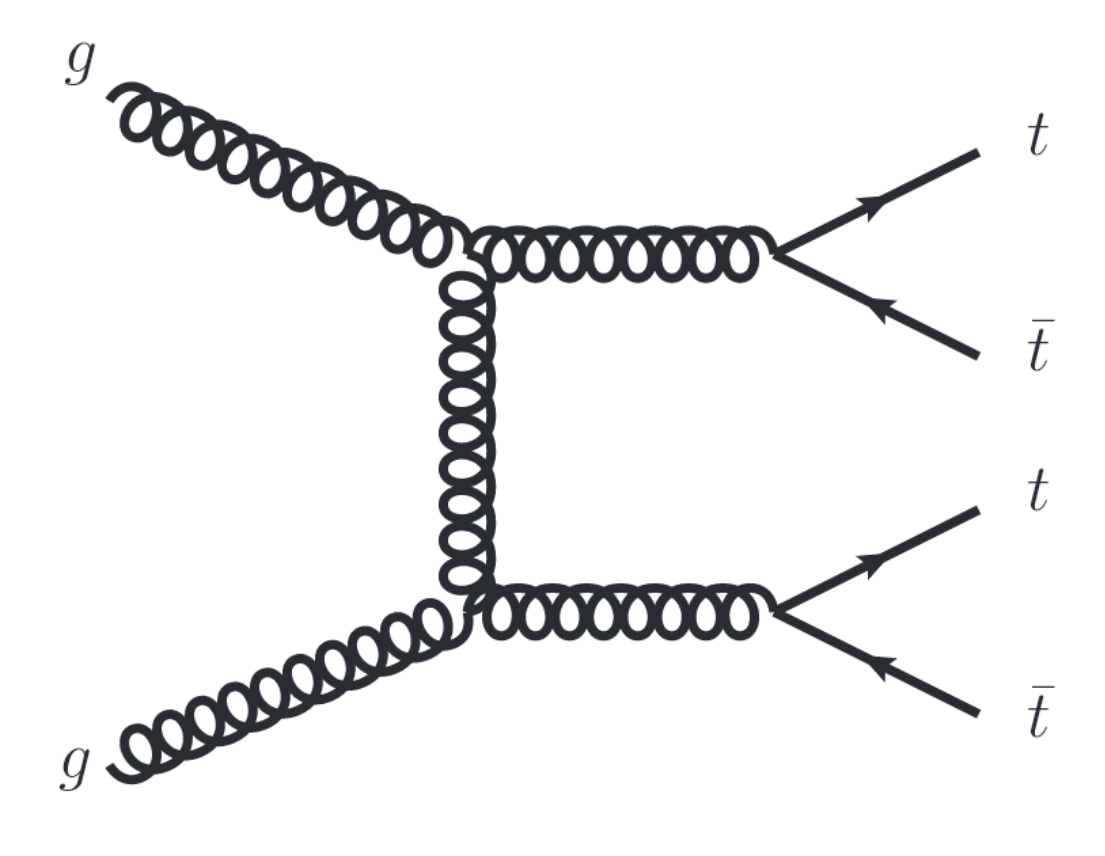}\hspace{1cm}
\includegraphics[width=0.27\linewidth]{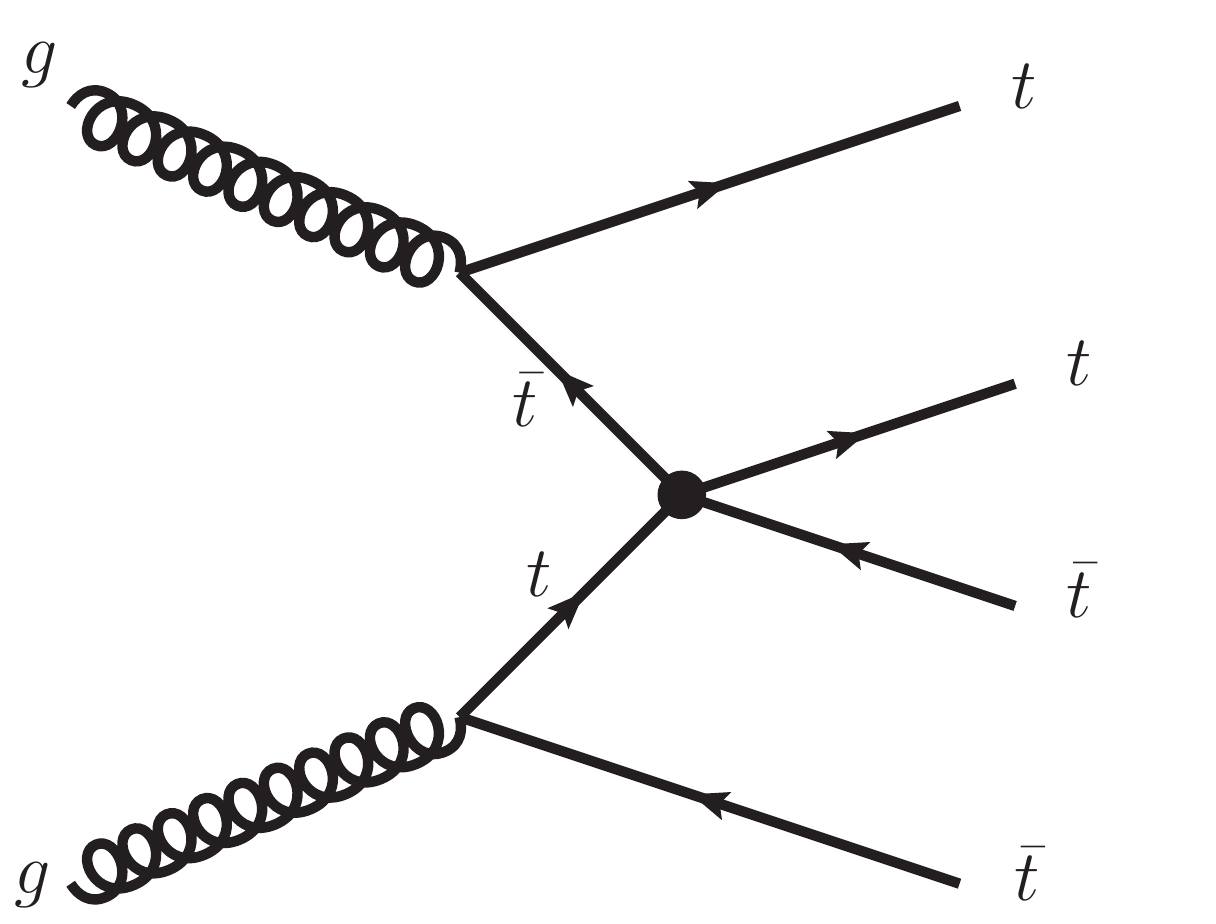}
\caption{Feynman diagrams showing four top quark production from the leading order SM contribution (left) and from contact interaction (right).}
\label{fig:diag_4tops}
\end{figure}

A direct search for physics beyond the Standard Model (BSM) is presented in a final state containing two isolated leptons (electrons or muons) with the same electric charge, $b$-jets and missing transverse momentum (\MET)~\cite{atlas_conf}. 
The analysis is performed with the dataset recorded in 2015 at the ATLAS experiment~\cite{ATLAS} at the LHC, corresponding to an integrated luminosity of 3.2 \fbi\ of $pp$ collisions at a center-of-mass energy of 13 TeV.
The search is sensitive to various signatures, in particular those containing four top quarks.
The Standard Model (SM) production of four top quarks is very small, but could be enhanced in various BSM scenarios.
At very high energies, this process can be described by an effective four-fermion contact interaction~\cite{CI}.
A Randall-Sundrum model with SM fields in the bulk~\cite{KK} is also considered in this search.
Feynman diagrams corresponding to the leading order SM contribution and to the contact interaction are shown in Figure~\ref{fig:diag_4tops}.
The search is also sensitive to the production of vector-like $T$, $B$, and $T_{5/3}$ quarks.

Several signal regions targeting the different scenarios are defined and described in Section~\ref{sec:signal}.
The estimation of the background contributions is described in Section~\ref{sec:bkg}.
The systematic uncertainties affecting the search are described in Section~\ref{sec:systs}, and the results are given in Section~\ref{sec:results}.

\section{Definition of the signal regions}
\label{sec:signal}
\begin{figure}[htb]
\centering
\includegraphics[width=0.3\linewidth]{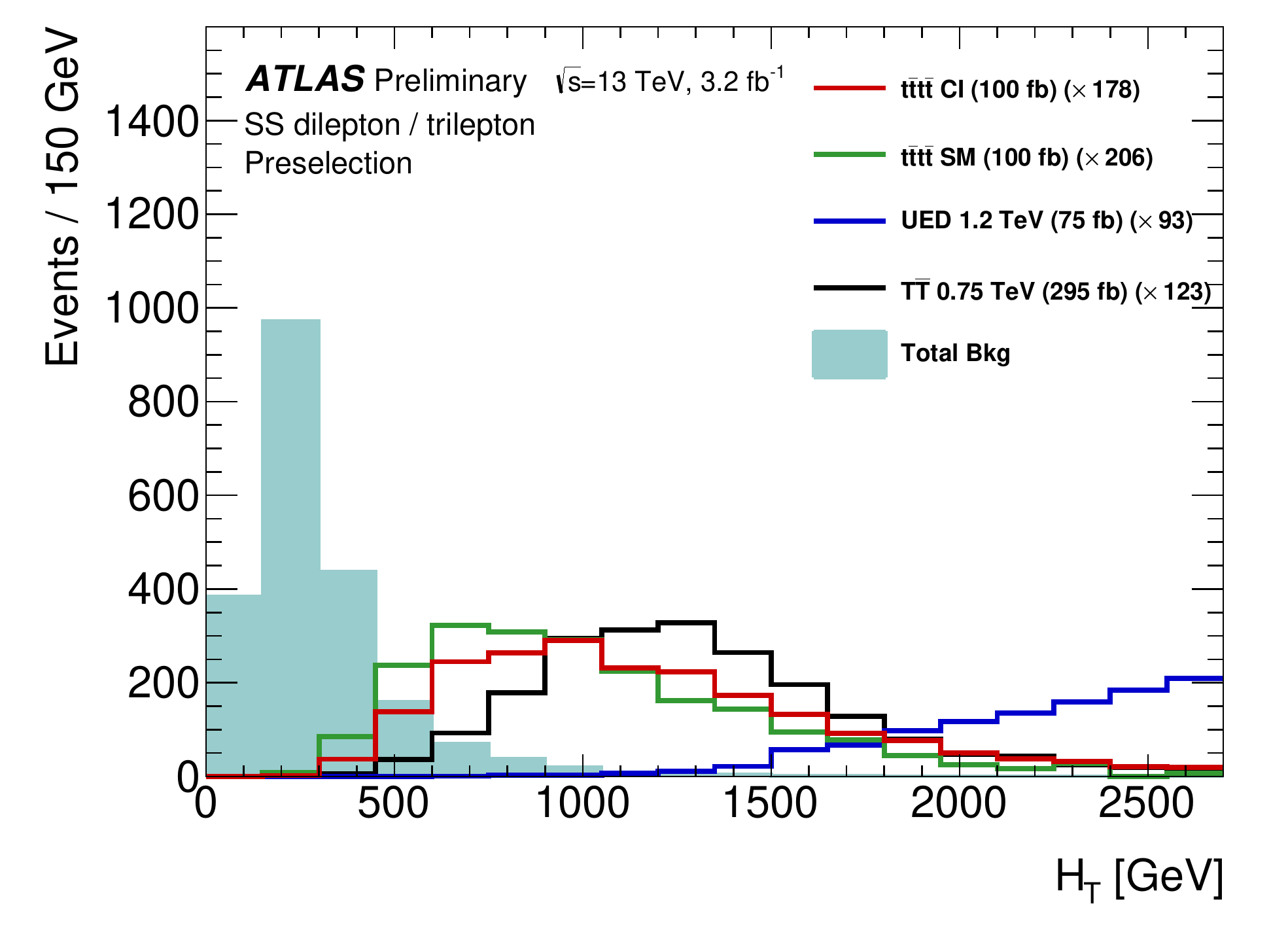}
\includegraphics[width=0.3\linewidth]{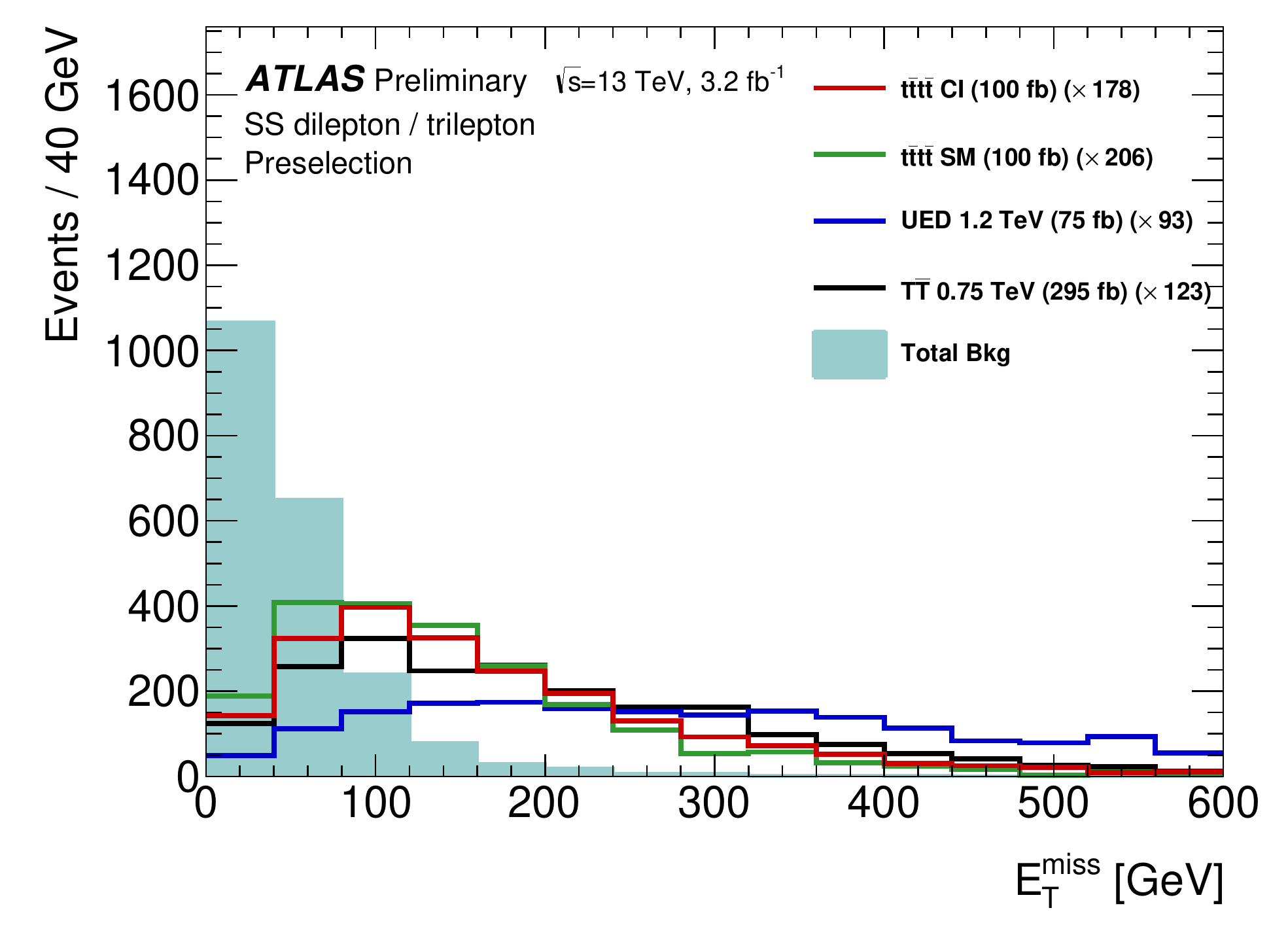}
\includegraphics[width=0.3\linewidth]{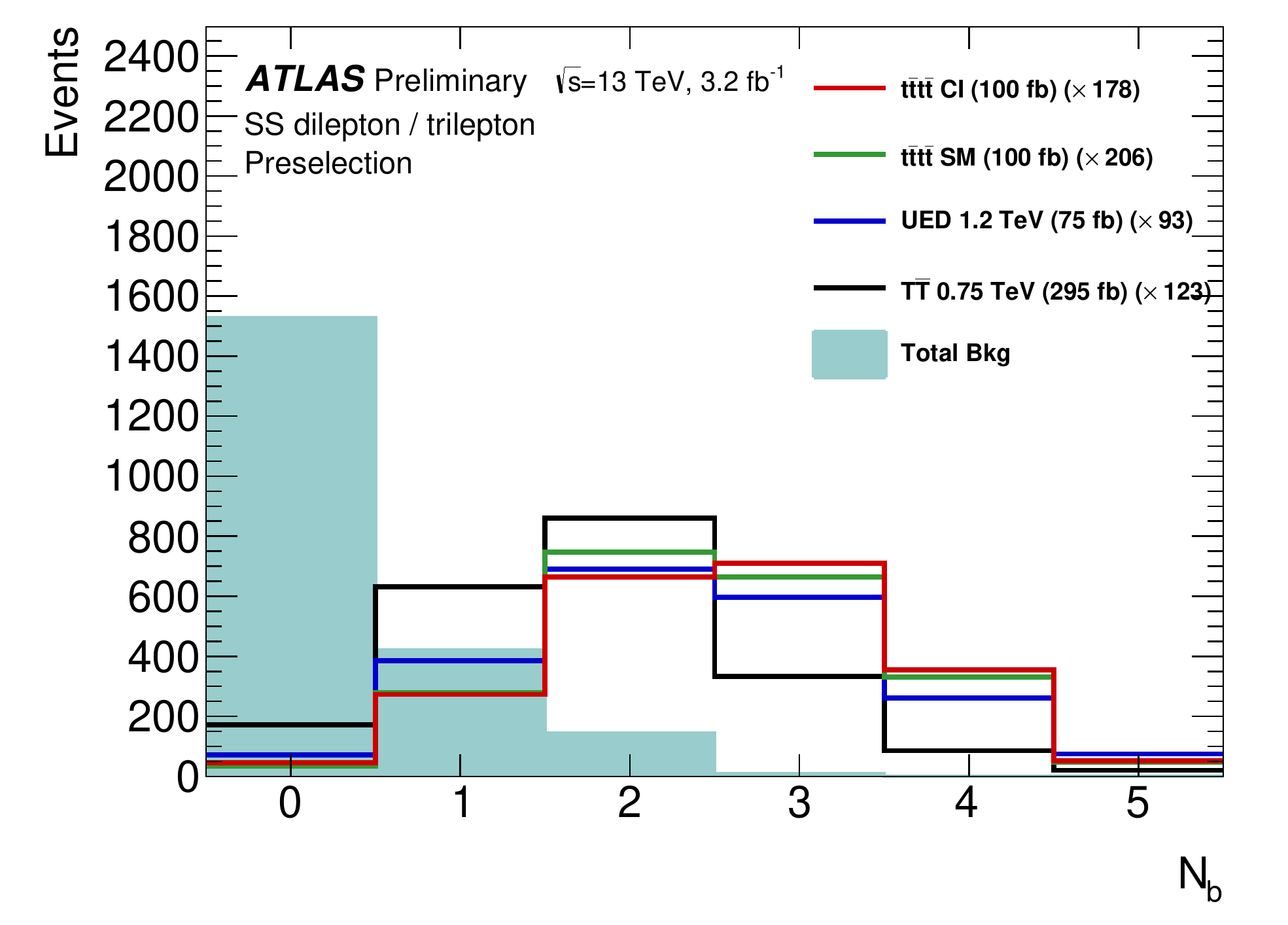}
\caption{Distributions of \HT\ (left), \MET\ (middle) and $N_b$ (right) for the background and different signals~\cite{atlas_conf}.}
\label{fig:distr_ht_met}
\end{figure}

This search is performed by defining eight signal regions sensitive to different BSM scenarios, and by comparing the expected number of events in the SM with the total number of events observed in data.
Preselected events contain at least two leptons, of which one pair has same electric charges.
The variables which are the most discriminant against the SM background and that allow to separate the different signals are: the number of jets containing $b$-hadrons ($N_b$), the \MET\ and the scalar sum of the transverse momenta of jets and leptons (\HT).
Distributions of these variables for the background and signals are presented in Figure~\ref{fig:distr_ht_met}.
The signal regions are presented in Table~\ref{tab:SR_def}.

\begin{table}[htb]
\begin{center}
\caption{Definition of the signal regions~\cite{atlas_conf}.}
\includegraphics[width=0.6\linewidth]{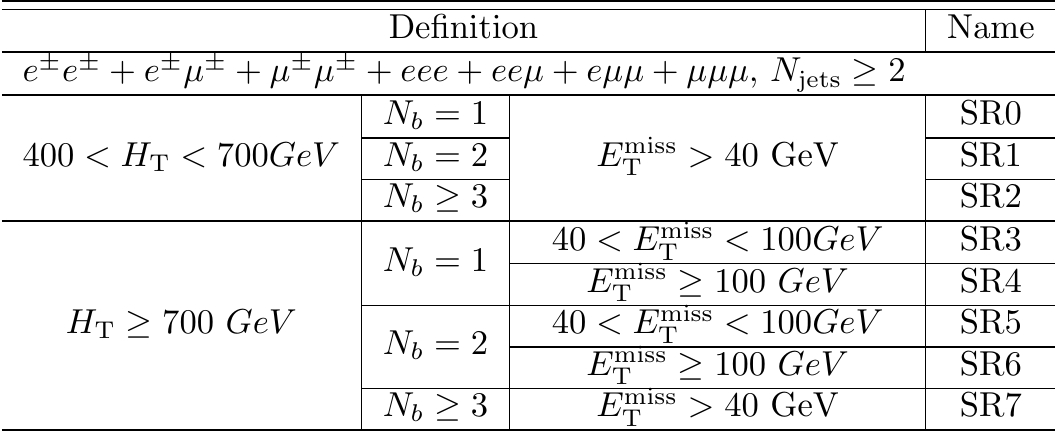}
\label{tab:SR_def}
\end{center}
\end{table}
\vspace{-1cm}

\section{Background estimation}
\label{sec:bkg}

\begin{figure}[htb]
\centering
\includegraphics[width=0.3\linewidth]{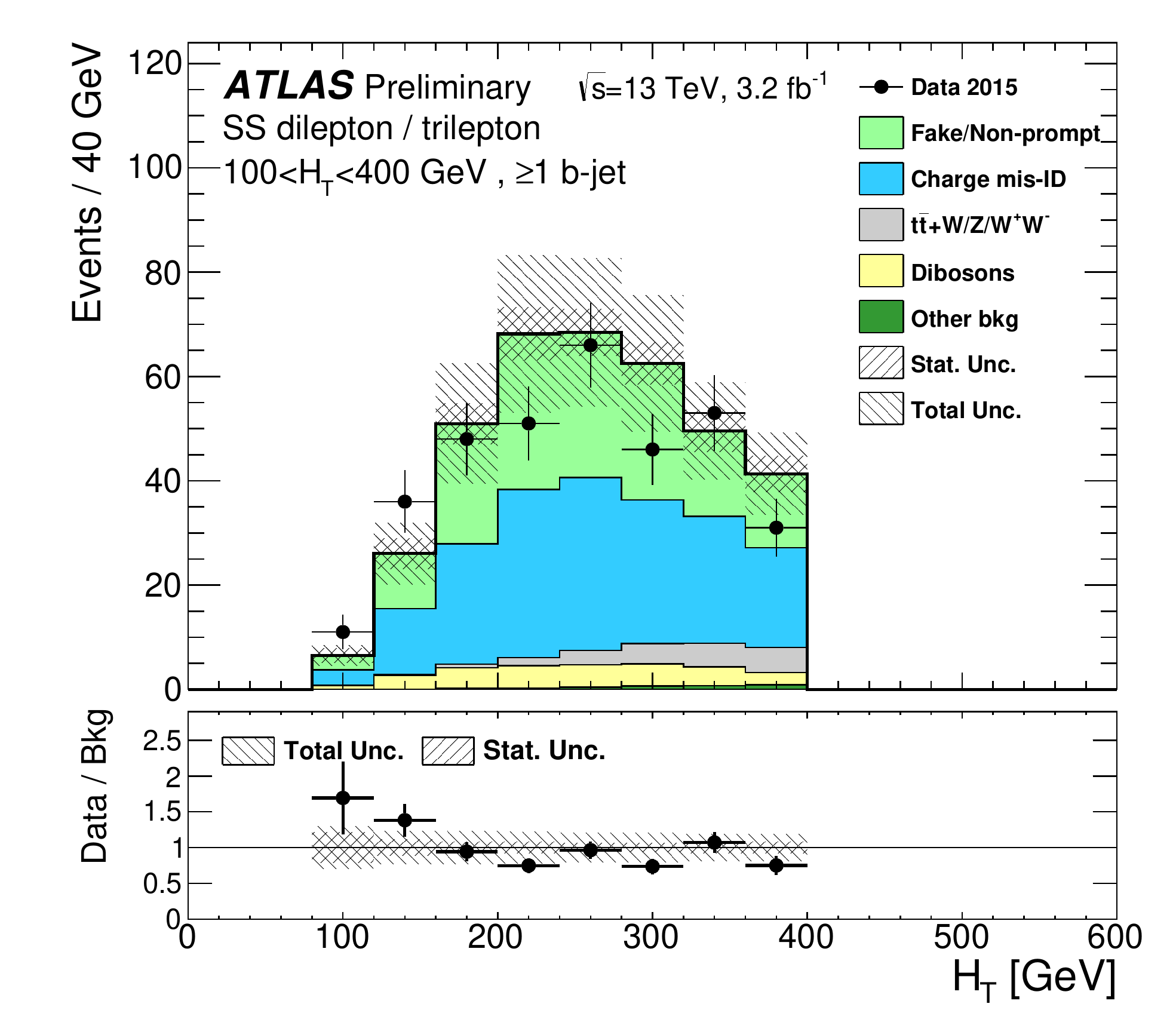}\hspace{1cm}
\includegraphics[width=0.3\linewidth]{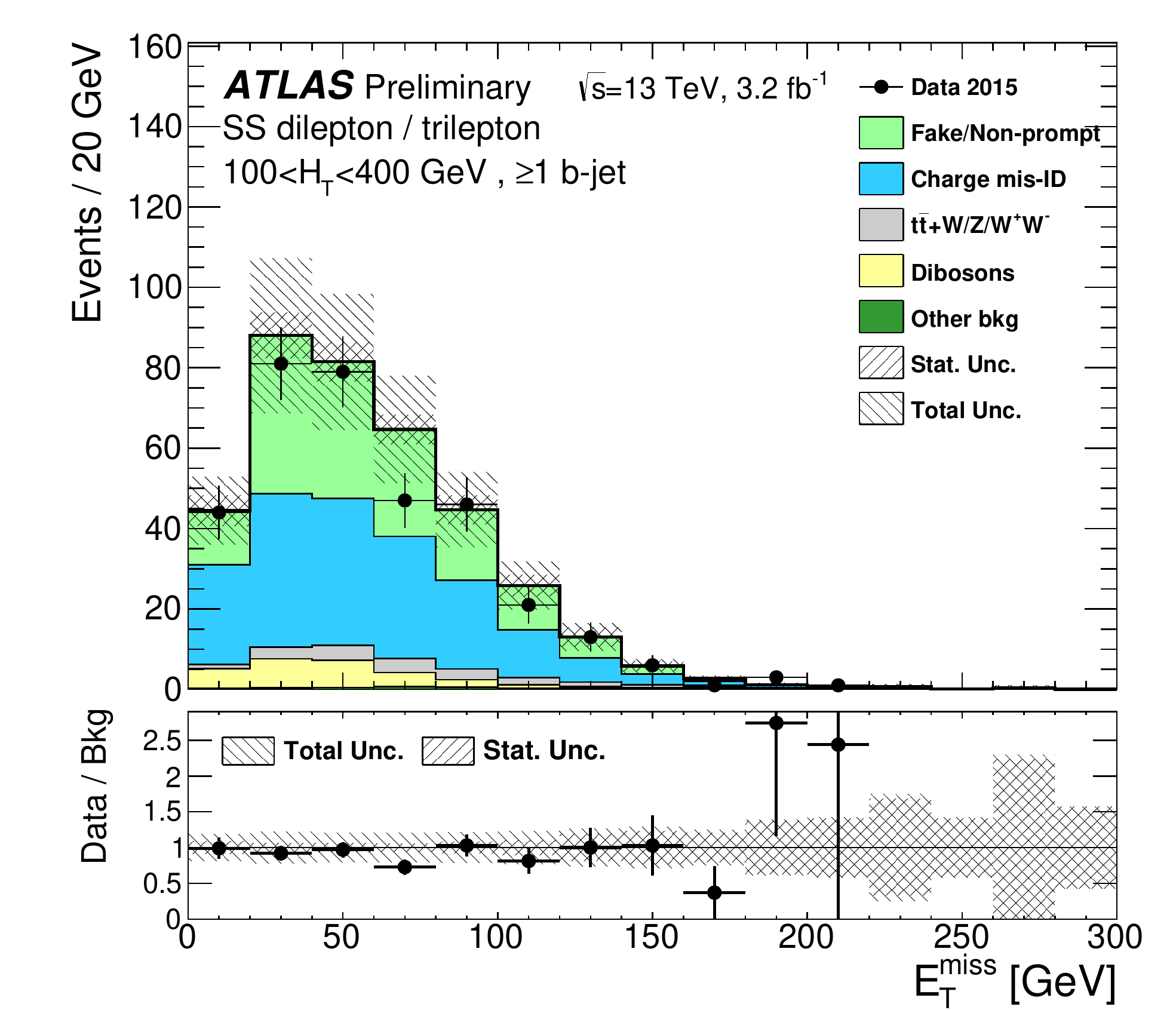}

\caption{Distributions of \HT\ (left) and \MET\ (right) in the control region with low \HT\ and at least one $b$-jet~\cite{atlas_conf}. The bottom panel of each plot shows the data-to-background ratio.}
\label{fig:dataMCcomp}
\end{figure}

The background is composed of three categories, each contributing roughly to one third of the total background contribution.
The first category corresponds to SM processes with two same-sign leptons in the final state. The main contributions are from the associated production of a $t\bar{t}$ pair and a gauge boson ($W$ or $Z$), diboson or triboson processes, as well as $t\bar{t}H$ and three top quark production.
They are estimated using Monte Carlo (MC) simulation.
The second category corresponds to events with so-called ``fakes'' or non-prompt leptons. Events from this category contain either a jet which has been reconstructed as an electron, or a lepton produced from a semi-leptonic decay of a $b$-hadron which passes the isolation criteria. This background is estimated using the matrix method applied on data~\cite{MM}.
The last category corresponds to events where the charge of an electron is misreconstructed. The contribution from this background is estimated using $Z \to ee$ events in a control region enriched in such events.
Some control regions, close but orthogonal to the signal regions are defined to check the estimation of the backgrounds.
Two examples are shown in Figure~\ref{fig:dataMCcomp}.

\section{Systematic uncertainties}
\label{sec:systs}
The effect of various systematic uncertainties are estimated on the background and signal yields.
The luminosity uncertainty is affecting both of them and is of 2.1\%. The detector modeling uncertainties are between 1 and 21\% on the signals, and 1 and 7\% on the backgrounds.
The main uncertainties on the backgrounds are on the estimation of the fake lepton background (54\%) and on the charge misidentification (25\%).
In both cases, the uncertainties are obtained by varying the control regions used in data and the MC models used to simulate real background contributions in these control regions. A large part from the uncertainty also comes from the limited available statistics of data.
The uncertainties on the cross-sections of the processes estimated from MC simulation are between 8 and 57\%.

\section{Results and conclusion}
\label{sec:results}
\begin{figure}[htb]
\centering
\includegraphics[width=0.5\linewidth]{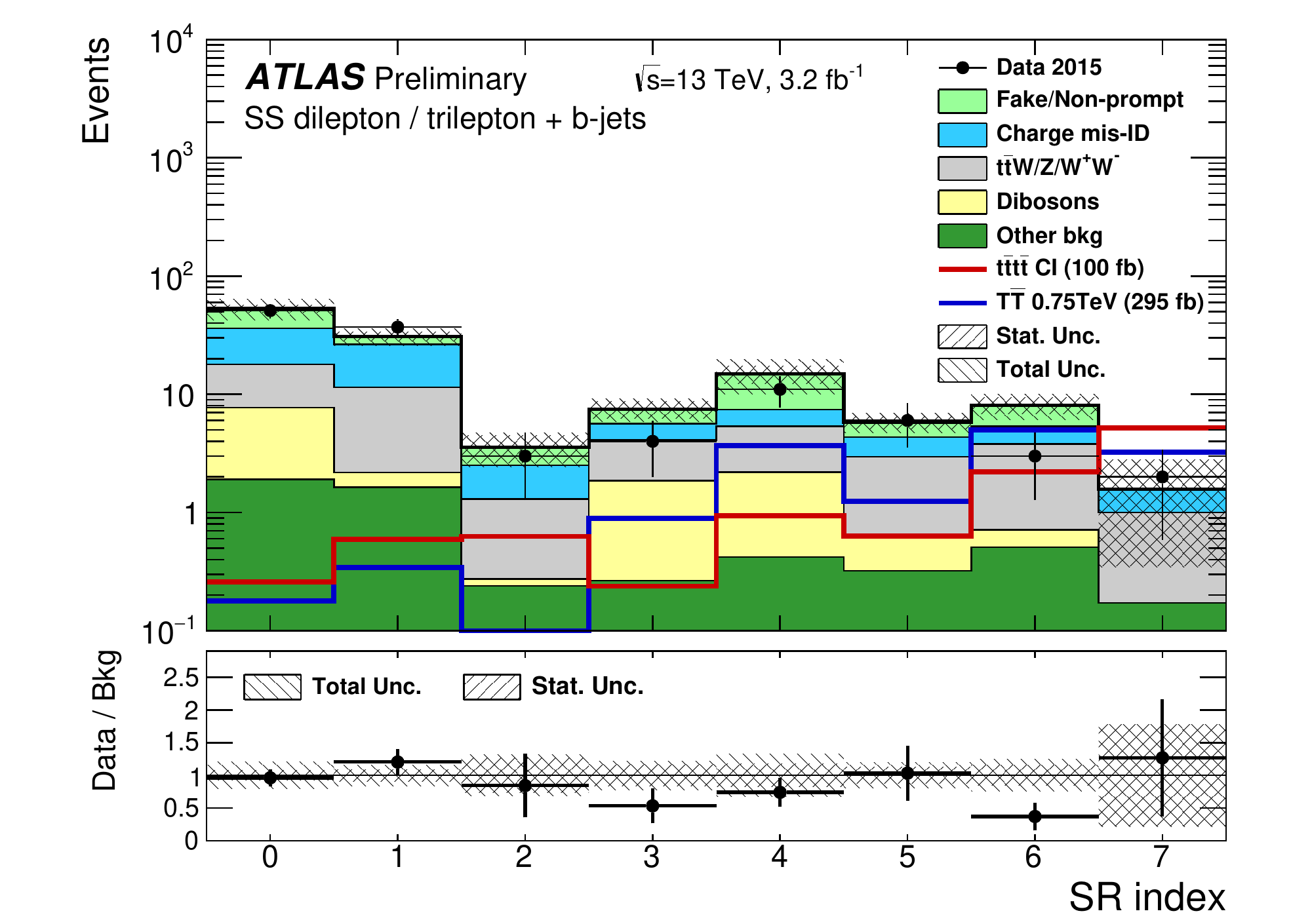}
\caption{Comparison between the observed and expected numbers of events in each signal region~\cite{atlas_conf}. The bottom panel shows the data-to-background ratio.}
\label{fig:SR_yields}
\end{figure}

After the background and systematic estimations, the observed number of events in each signal region is compared to the expected one.
The comparison is shown in Figure~\ref{fig:SR_yields}.

\begin{figure}[htb]
\centering
\includegraphics[width=0.45\linewidth]{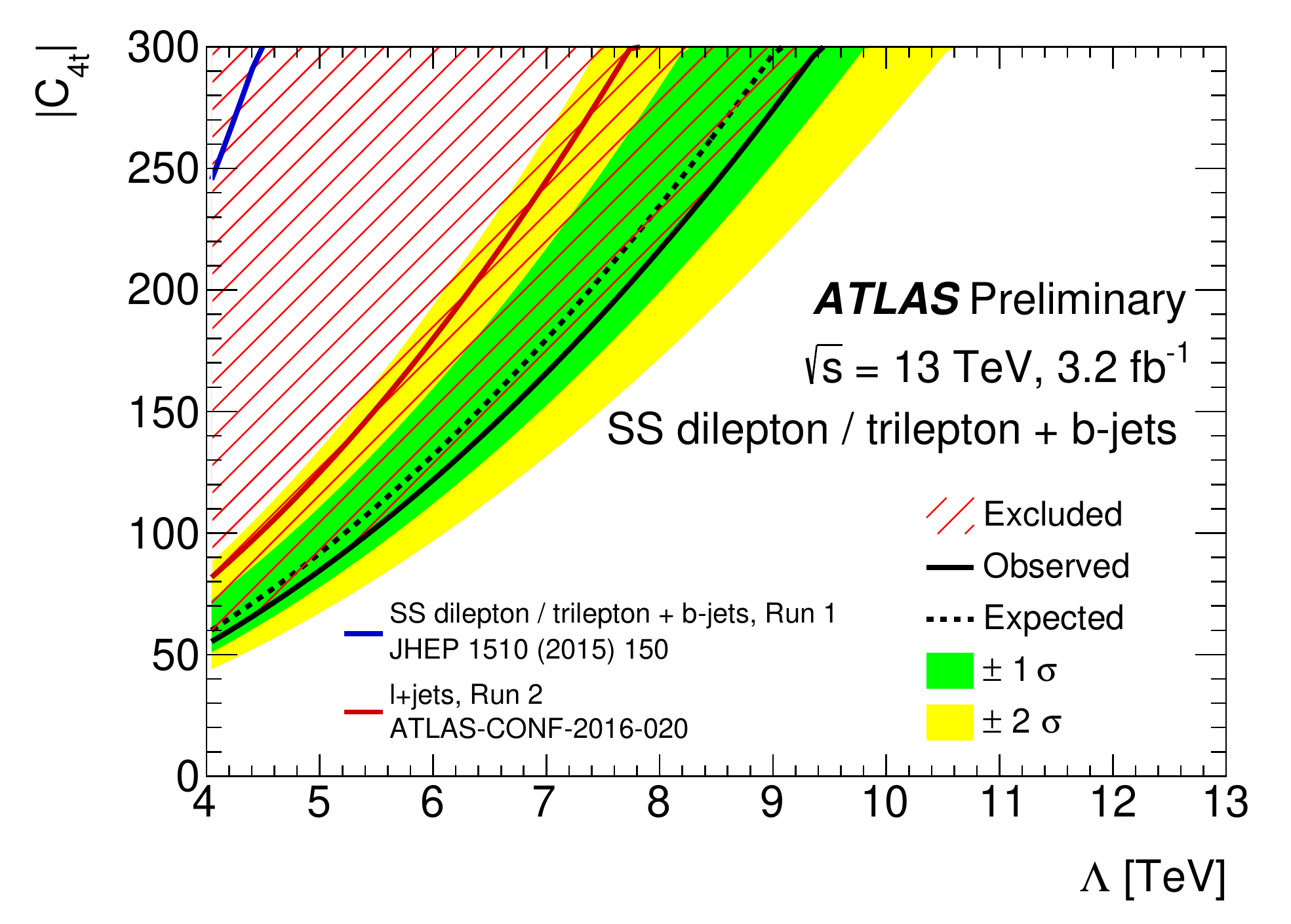}
\includegraphics[width=0.45\linewidth]{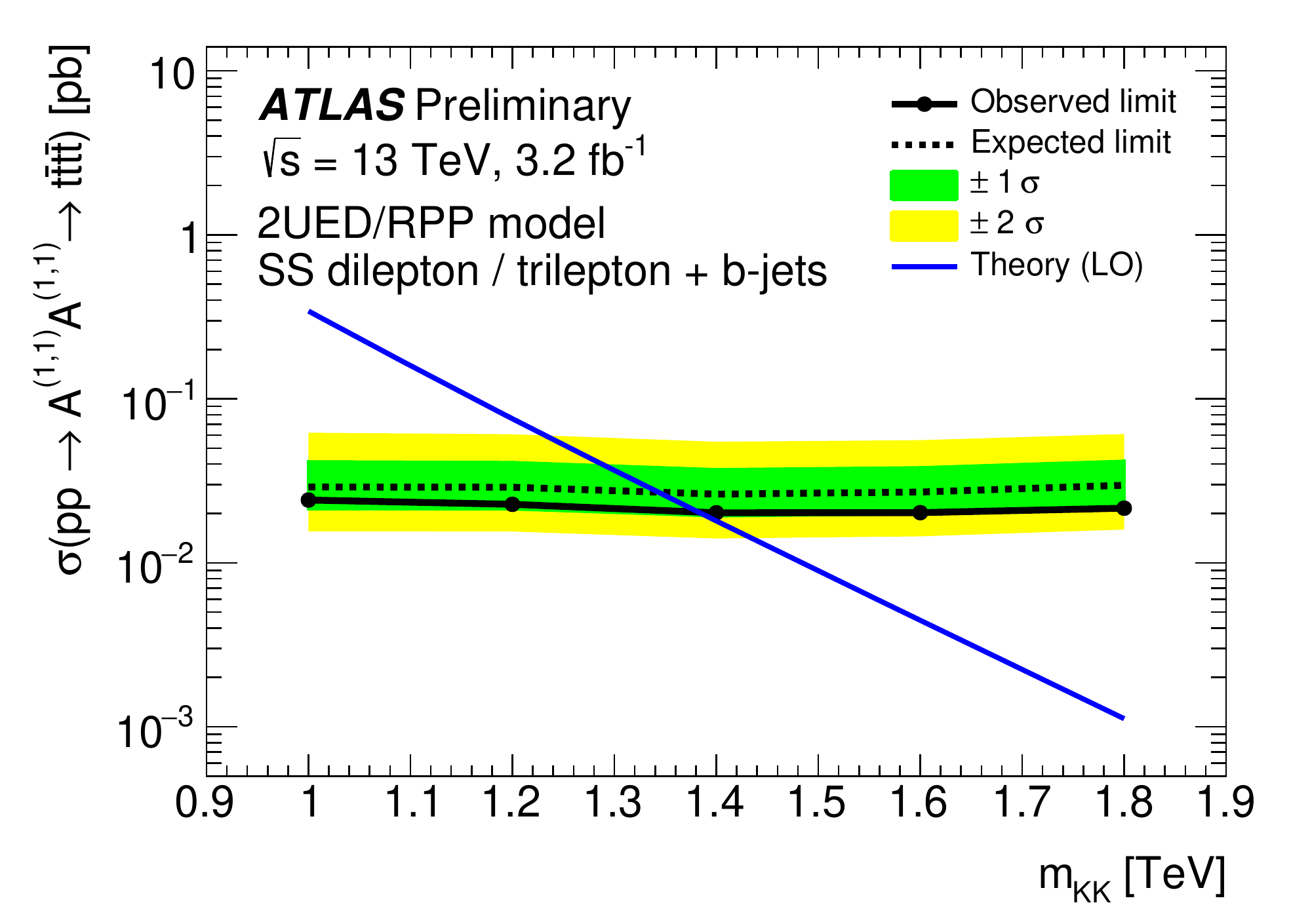}
\caption{Limits on the coupling and energy scale of the contact interaction model (left) and on the mass of a Kaluza-Klein resonance leading to four top quark production (right)~\cite{atlas_conf}.} 
\label{fig:limits}
\end{figure}

No excess is observed in data, and some limits are set on the various BSM physics models using a profile log-likelihood ratio.
For the four top quark production, limits are set on the cross-section. For SM kinematics, the limit at 95\% confidence level is 95~fb (with an expected limit of 107~fb). This corresponds to approximately ten times the SM cross-section. For the contact interaction model, the limit is 67~fb (79~fb expected). Some limits are also set on $C_{4t}$ and $\Lambda$, the coupling and energy scale used to parametrize the contact interaction model. These are shown in Figure~\ref{fig:limits} with the limits derived on the mass of a Kaluza-Klein resonance.

\begin{thebibliography}{99}


\bibitem{atlas_conf} ATLAS Collaboration, ATLAS-CONF-2016-032 (2016)\\ https://cds.cern.ch/record/2161545
\bibitem{ATLAS} ATLAS Collaboration, JINST \textbf{3} (2008) S08003
\bibitem{KK} M. Guchait, F. Mahmoudi, and K. Sridhar, Phys. Lett. B \textbf{666} (2008) 347-351 
\bibitem{CI} C. Degrande et al, JHEP \textbf{03} (2011) 125
\bibitem{MM} D0 Collaboration, Phys. Rev. D \textbf{61} (2000) 072001


\end{thebibliography}
\end{document}